# Symmetry and symmetry-breaking in soil pores and climate change mitigation: What fractal geometry can tell us?

Abhijeet Das[1]

[1]Department of Bioengineering, Indian Institute of Science, Bangalore – 560012, India

Email – abhijeetdas@iisc.ac.in

Humans are searching for signs of extraterrestrial life by analyzing the Martian soil. This raises an interesting question. Why soil was explicitly chosen? The answer is that it attained this distinct status as it supports all life forms on our own planet from the ground below. Interestingly, despite its significance, we knew little about it.

It was reported that the spatial clustering of soil and microorganisms provides soil with its characteristic aggregated structure[1]. Young and Crawford[2] in an issue of Science reported this characteristic structure responsible for determining the physical properties of pores consequently, the diffusion rate of gases into and out of the soil. They also emphasized the utilization of fractal geometry in the investigation of soil structure and related consequences. Recently, it was argued that all soil characteristics and properties boil down to its structure, quantified by the architectural, energetics, and chemical parameters. They further showed that soil possesses a dual nature arising from the interlinking between the soil's solid phase and pores and termed the resultant groupings aggregates[3]. However, pores alone were reported to sufficiently describe air (or gas) permeability from the soil. Moreover, initial pore morphology was argued to be a prominent factor for the determination of the effect on pore network architecture and subsequently, microbial activity, water transport/retention, etc. by the freezing-warming cycle in permafrost aggregates of arctic landscapes[4].

Given the self-organized hierarchical organization of soil, this bottom-up organized complex system can be considered to be constituted of a complex system of pores, giving rise to interesting properties in pore size, shape, localization, connectivity, and consequently, emergent properties like permeability, and diffusivity. The aforementioned highlights the prominent role of pores in greenhouse gas emissions and exchange from soil[5,6].

Captivatingly, the concept of symmetry and symmetry-breaking has been lately introduced where, combinatorial, geometric, and functional symmetries were argued to comprehensively describe the complexity of a system[7]. We shall discuss in this perspective, the possible application of symmetry for investigating the structure-properties-function relation in soil-pore. Nonetheless, it should be realized that though individual symmetry components can be studied and described independently, nature utilizes them in combination.

Combinatorial symmetry refers to a diverse range of sizes, possible combinations, and interconnectedness of pores with varying degrees of complexity and connectivity. A possible framework to utilize in these investigations is fractal geometry where mono-fractal parameters like (i) fractal dimension quantify the degree of statistical self-similarity and correlate the measured metrics (pore shape and size) complexity and scale of investigation and (ii) lacunarity correlate the spatial distribution of pores to the measuring scale along with their rotational/translational invariance. In addition, the Hurst exponent can be utilized to investigate the memory effect in soil properties where memory would signify information related to factors and processes which contributed to soil formation. Furthermore, multi-fractal measures like multifractality strength, and fractal dimension for individual Hölder exponent can reflect on different aspects of complexity in distributions of pores or pore-size diversity. Additional fractal parameters like succolarity and Shannon entropy can shed light on tortuosity, connectivity, and randomness in soil-pore distribution, respectively. The robustness and resilience of complex pore networks against perturbations from immediate and distant factors can be studied using additional approaches as discussed here[8].

Oliveira *et al.*[9] studied the effect of wetting-drying cycles on pore network complexity with the secondary forest as control and different management practices (conventional, minimum, and no tillage) utilizing three-dimensional multi-fractal analysis implemented on X-ray microtomography images. Lacunarity analysis indicated an increase (decrease) in the dispersion of soil-pore clusters with sequential periods of wetting-drying in secondary forest (minimum and no tillage) whereas, no substantial heterogeneity in pore distribution was reported for conventional tillage practice. In multi-fractal analysis, the presence of multifractality is generally confirmed from the generalized or Renyi dimension following the trend; capacity dimension ($D_0$) > information dimension ($D_1$) > correlation dimension ($D_2$) and/or from the non-linear decrement in the mass exponent (or singularity strength) with moment[10,11]. In the reported study, the existence of multifractal behavior in pore distribution was confirmed by the generalized dimension. In addition, an increase (decrease) in soil-pore randomness for conventional forestry following wetting-drying cycles was realized from the Shannon entropy. However, no clear implication of this parameter was found for conventional and minimum tillage practices.

In regard to geometric (including topology, morphology, and topography) symmetry, applying techniques or measures from fractal geometry is a possible way to highlight the architectural complexity and/or heterogeneity in pore space. The parameters representing the geometry of soil include shape, size, orientation, etc. where, the shape and orientation of pores specifically determine soil's morphological properties. Pires *et al.*[12] studied the effect of conventional and no tillage practices on soil structural variations with the increase in depth where they emphasized the micro-morphological and -topographical characteristics using two-and three-dimensional image analysis, respectively. They observed the prominence of similar pore structures (shape, size, orientation) for the lower layer (0-10 *cm*) with an increase in small and medium pore size for both practices. However, a greater contribution of large complex pores was realized in the case of land with no tillage practices. In addition, enhancement in pore shape irregularity between 10-30 *cm* depth, although with better connectivity, was observed from topographical analysis. The aforementioned characteristics of pore structure or architecture are a reminiscence of fractal structures implying the significance of fractal parameters in their description.

Functional symmetry is related to the effect on pore features from immediate factors (soil type and chemical properties, land practices) and distant factors (temperature, humidity). We have seen the effects of land practices like tillage in previous sections subsequently, will only focus on the remaining factors, especially concerning wetlands and barren lands. Wetlands have augmented moisture content and water-saturated soil layers which contributes to unique pore structures. The soil possesses both macro- and micro-pores in high proportion where the former is formed from the channels and fissures created by vegetation. In contrast, barren land characterized by low moisture content has only a high micropore proportion however, a reduction in pore connectivity can be realized from the accumulation of salts and minerals owing to lack of vegetation.

Soil pH affects nutrient availability and microbial activity, impacting soil structure and pore networks. Optimal pH fosters stable aggregates and interconnected pores, seen also in wetlands with low carbon (C)-to-nitrogen (N) ratios. However, variations in pH and C/N, influenced by organic matter and vegetation, can alter pore traits. Conversely, highly alkaline soil disperses particles and reduce pore connectivity, with low pH forming crusty soil. In addition, changes in C/N due to nutrient limitations affect pore characteristics significantly.

Humidity and temperature are critical factors as they highlight oxygen availability and organic matter decomposition along with the activity of microorganisms in the soil, respectively. Optimal humidity and temperature can result in pore connectivity augmentation, however, excessive increase in humidity and temperature can significantly change the pore characteristics in wetlands whereas, low (high) humidity (temperature) can lead to soil desiccation and crust formation with reduced connectivity between pores[5,6].

Although, we only discussed the possibilities in wetlands and barren lands the observation of pore features can be extended to forestland, grassland, croplands, and permafrost, respectively in the context of agricultural practices (fertilizer usage, tillage, following, and irrigation) and wetting-drying/freezing-thaw cycles.

This perspective only aimed to highlight the possible applicability of fractal geometry in investigating the discussed symmetry and their breakings in the case of soil pore features. This could in turn shed light on the microbial activity, diffusion of gases, water transport and retention, and air permeability in a diverse range of soil and subsequently, can aid in better understanding the emission and exchange of greenhouse gases ($CO_2$ and $N_2O$ emissions and $CH_4$ uptake) consequently, climate change mitigations. In addition, it can aid in understanding the self-organization of soil facilitated by solid and pore phase aggregates and their microenvironment.